\documentclass[12pt]{article}

\newcommand{\nc}{\newcommand}
\nc{\bib}{\bibitem}
\nc{\al}{\alpha}
\nc{\g}{\gamma}
\nc{\G}{\Gamma}
\nc{\D}{\Delta}
\nc{\eps}{\epsilon}
\nc{\om}{\omega}
\nc{\la}{\lambda}
\nc{\La}{\Lambda}
\nc{\var}{\varphi}
\nc{\pa}{\partial}
\nc{\nn}{\nonumber \\ }
\nc{\hf}{\frac{1}{2}}         
\nc{\dz}{\frac{dz}{2\pi i}}
\nc{\bin}[2]{\left (\begin{array}{c} {#1}\\ {#2} \end{array}\right )}
\nc{\ben}{\begin{equation}}
\nc{\een}{\end{equation}}
\nc{\bea}{\begin{eqnarray}}
\nc{\eea}{\end{eqnarray}}
\nc{\bra}[1]{\langle {#1}|}
\nc{\ket}[1]{|{#1}\rangle}
\newcommand{\Z}{\mbox{$Z\hspace{-2mm}Z$}}
\nc{\C}{\mbox{\hspace{1.24mm}\rule{0.2mm}{2.5mm}\hspace{-2.7mm} C}}
\nc{\Nat}{\mbox{\hspace{.04mm}\rule{0.2mm}{2.8mm}\hspace{-1.5mm} N}}

\renewcommand{\thefootnote}{\fnsymbol{footnote}}
\nc{\gb}{\bar{g}}
\nc{\gt}{\tilde{g}}
\nc{\ut}{\tilde{U}}
\nc{\zb}{\bar{z}}
\nc{\ad}{\rm{Ad}}
\begin{document}

\topmargin -5mm
\oddsidemargin 5mm

\begin{titlepage}
\setcounter{page}{0}
\vspace{8mm}
\begin{center}
{\huge Maximally symmetric D-branes}\\[.2cm]
{\huge in gauged WZW models}

\vspace{15mm}

{\large Takahiro Kubota}\footnote{kubota@het.phys.sci.osaka-u.ac.jp}\\[.2cm]
{\em Department of Physics, Graduate School of Science, 
Osaka University, Toyonaka, Osaka 560-0043, Japan}\\[.5cm]
{\large J{\o}rgen Rasmussen}\footnote{rasmusse@crm.umontreal.ca}\\[.2cm]
{\em CRM, Universit\'e de Montr\'eal, Case postale 6128, 
succursale centre-ville, Montr\'eal, Qu\'ebec, Canada H3C 3J7}\\[.5cm] 
{\large Mark A. Walton}\footnote{walton@uleth.ca} and 
{\large Jian-Ge Zhou}\footnote{jiange.zhou@uleth.ca}
\\[.2cm]
{\em Physics Department, University of Lethbridge,
Lethbridge, Alberta, Canada T1K 3M4}

\end{center}

\vspace{10mm}
\centerline{{\bf{Abstract}}}
\vskip.4cm
\noindent
Gluing conditions are proposed to characterize the D-branes
in gauged WZW models. From them the boundary conditions for 
the group-valued and the subgroup-valued fields are determined. 
We construct a gauged WZW action for open strings that
coincides classically with those written previously, when the 
gluing conditions are imposed.

\end{titlepage}
\newpage
\renewcommand{\thefootnote}{\arabic{footnote}}
\setcounter{footnote}{0}

\section{Introduction}

D-branes on group manifolds are characterized by the gluing condition
$J=-R\bar{J}$ at the boundary \cite{AS}-\cite{ssd}.
It has been shown that the maximally symmetric gluing condition
$J=-\bar{J}$ (in the closed string picture) constrains  
the end points of open strings to conjugacy classes of 
the group manifold \cite{AS}.
To make the Wess-Zumino-Witten (WZW) action for open strings  
well-defined, the position of the conjugacy classes has to be quantized
-- in perfect agreement with the Cardy boundary states. 
Related aspects of D-branes on group manifolds have been discussed
in \cite{Gaw1}--\cite{ssi} using various approaches.

D-branes on coset models have been studied more 
recently \cite{MMS}--\cite{hi}.
It was shown how the D-branes can be given a
geometrical interpretation from the point of view of the gauged WZW model.
In \cite{Gaw2,ES}, more general boundary conditions were proposed
for the group-valued field $g$ in the gauged WZW model, and
an algebraic classification of Cardy boundary states on a $G/H$ coset
conformal field theory (CFT) has been realized geometrically on the
gauged WZW model.

In the $G/H$ gauged WZW model, the boundary value of the group-valued
field $g\in G$ was suggested to be constrained to a product of  
two conjugacy classes, one for $G$ and the other for $H$ \cite{Gaw2,ES}. 
This result was reached essentially by demanding that the vector (diagonal) 
gauge symmetries of $g$ in the 
bulk are conserved when the boundary conditions are imposed.
The boundary condition for the gauge field $A$ was mentioned in \cite{ES}, 
but there $g$ is assumed to be in 
a single conjugacy class of $G$ instead of in the product of
two conjugacy classes. It is interesting then to see how one
can give a consistent description of the boundary conditions
for the group-valued field $g$ and the gauge field $A$ (or equivalently, for 
$g$ and fields $U,\ut$).

As already mentioned, in the WZW model the maximally symmetric D-branes
are characterized by the gluing condition $J=-\bar J$. Then 
the end points of open strings are restricted to the conjugacy
classes of the group $G$ \cite{AS}. Based on this experience 
with the boundary WZW model, it is natural to ask 
if there are similar gluing conditions 
for the gauged WZW model that encode the boundary
conditions for $g,U,\ut$. That is, can the corresponding D-branes 
be characterized by gluing conditions?

Here we propose such gluing conditions for the gauged WZW model 
that resemble those for the ordinary WZW model. 
They enable us to characterize the D-branes
explicitly, and to determine the boundary conditions for all fields in the
coset model.
Our results pertain to the maximally symmetric D-branes in the gauged
WZW model. We recover the boundary conditions on $g$ \cite{Gaw2,ES},
and also extend the result to cover gluing
conditions twisted by inner automorphisms
of $G$ reducing to $H$. We find that in the product of
the twisted conjugacy classes, the twistings cancel, and we are
left with the same product of (untwisted) conjugacy classes 
appropriate to the untwisted case. 

Furthermore, from our gluing conditions, 
we construct the gauged WZW action for 
open strings. The new feature is that the two-forms $\om$ depend on the
boundary value of one of the $H$-valued fields $U,\ut$ ($U$, say).
When rewriting our action in terms of the gauge field $A$, 
we find that the resulting action coincides classically with that
in \cite{Gaw2,ES}. This is possible because 
the $U$-dependent terms from different sources cancel
in a nontrivial way; 
this indicates that our gluing conditions indeed describe 
D-branes in the gauged WZW model. 

The layout of this note is as follows. 
In Section 2, we introduce the gluing conditions  $J=-\bar J$
and $K=-\bar K$ in the gauged WZW model, and obtain the
boundary conditions for fields $g, U,\ut$. We then consider
twisted gluing conditions, and find that
inner automorphisms of $H$ do not change 
the boundary restrictions of $g$ --- only $U,\ut$ are affected. 
In Section 3, we construct the
gauged WZW action for open strings from our gluing conditions,
and compare it with that in \cite{Gaw2,ES}. 
A short summary is presented in Section 4 as our conclusion.  

\section{Gluing conditions in the gauged WZW model}

The level-$k$ WZW action for closed strings on a compact, connected, 
simply connected,  Lie group $G$ may be written
\bea
 S^G(g)&=&\frac{k}{4\pi}\left(\int_\Sigma d^2z\,L^{\rm kin}(g)
+\int_B\chi(g)\right)
\label{SG}
\eea
where $L^{\rm kin}(g)=tr(\pa_z g\pa_{\zb}g^{-1})$, $\chi(g)=
\frac{1}{3}tr(dgg^{-1})^3$, and  $B$ is a three-dimensional manifold
bounded by $\Sigma$. The chiral currents are
\ben
{\cal J}_z=-\pa_z g g^{-1},\ \ \ \ \bar{{\cal J}}_{\zb}=g^{-1}\pa_{\zb}g
\label{J}
\een
corresponding to the ``global'' symmetries $g(z,\bar z)\rightarrow L(z) 
g(z,\bar z)$, $g(z,\bar z)\rightarrow g(z,\bar z)R(\bar z)^{-1}$, 
respectively. 

One may gauge the vector (diagonal) subgroup of this 
``$G_L(z)\times G_R(\zb)$'' symmetry. The resulting local symmetry 
transformation 
is then  
$g(z,\zb)\rightarrow m(z,\zb)g(z,\zb)m(z,\zb)^{-1}$, with 
$m$ an element of the subgroup $H\subset G$. Introduce the 
``gauge fields'' $U,\ut\in H$, transforming as $U\rightarrow mU$, 
$\ut\rightarrow m\ut$. A gauge invariant action is 
\ben
 S^{G/H}=S^G(U^{-1}g\ut)-S^H(U^{-1}\ut)
\label{SGHU}
\een
Here $S^H$ is defined as in (\ref{SG}) except that the trace used 
$tr_H\{\cdot\}$ is that 
appropriate to the Lie algebra ${\rm Lie}\,H$  
of $H$, and the level is $k_H$. Let 
$t_H$ denote an element of ${\rm Lie}\,H$, and let $\epsilon$ 
be the embedding of ${\rm Lie}\,H$ in ${\rm Lie}\,G$, so that $\epsilon(t_H)$ 
is the element of the subalgebra ${\rm Lie}\,H\subset {\rm Lie}\,G$ 
corresponding to $t_H$. Now $tr_G(\epsilon(t_H))= 
\iota\, tr_H(t_H)$, where the subscript $G$ is included for clarity, and 
$\iota$ is the index of 
the embedding. Therefore, 
by the construction of (\ref{SGHU}), 
$k_H=\iota k$.   

Making use of the Polyakov-Wiegmann identities 
\bea
&&L^{\rm kin}(g_{1}g_{2})=L^{\rm kin}(g_{1})+L^{\rm kin}(g_{2})-tr(g_{1}^{-1}
\pa_z g_{1}\pa_{\zb} g_{2}g_{2}^{-1}+g_{1}^{-1}\pa_{\zb}g_{1}
\pa_z g_{2} g_{2}^{-1})\nn
&&\chi (g_{1}g_{2})=\chi (g_{1})+\chi (g_{2})-d\,tr(g_{1}^{-1}dg_{1}\,dg_{2}
g_{2}^{-1})
\eea
and defining 
\ben
 A_z=\pa_z \ut\ut^{-1},\ \ \ \ A_{\bar{z}}=\pa_{\bar{z}}UU^{-1}
\label{A}
\een
results in the gauged 
WZW action \cite{gk,kps}
\ben
 S^{G/H}=S^G+\frac{k}{2\pi}\int_\Sigma d^2z\,tr\left\{A_{\zb} \pa_z gg^{-1}
  -A_z g^{-1}\pa_{\zb} g+A_{\zb} gA_z g^{-1}-A_z A_{\zb} \right\}
\label{SA}
\een
Note that the vector fields $A_z, A_{\bar z}$ take values in  
${\rm Lie}\,H$.

Working backwards then, 
the gauged WZW action (\ref{SA}) may be rewritten as a coset 
action \cite{gk,kps}
\ben
 S^{G/H}=S^G(\gt)-S^H(h)
\label{SGH}
\een
with
\ben
 \gt=U^{-1}g\ut\in G,\ \ \ \ h=U^{-1}\ut\in H,\ \ \ \ g\in G,\ \ \ \ 
   U,\ut\in H
\label{gth}
\een

Each WZW action in the gauged WZW action $S^{G/H}=S^G(\gt)-S^H(h)$ 
is well defined for the fields $\gt$ and $h$ \cite{gk},
and the associated currents are
\bea
 &&J=-\pa_z\gt\gt^{-1},\ \ \ \ \bar{J}=\gt^{-1}\pa_{\bar{z}}\gt\nn
 &&K=\pa_z hh^{-1},\ \ \ \ \bar{K}=-h^{-1}\pa_{\bar{z}}h
\label{JK}
\eea
$J$ and $K$ are associated with the groups $G$ and 
$H$, respectively, and the sign difference on the currents 
stems from the minus sign in front of $S^H$ in (\ref{SGH}).
The two currents are not independent \cite{kps}, however,  
since there is a first class constraint. Recall that the 
Lie algebra embedding is denoted 
$\epsilon:\, {\rm Lie}\,H\rightarrow {\rm Lie}\,G$. 
The components of 
$J$ and $\epsilon(-K)$ in the Lie subalgebra $H\subset G$ are 
clearly related.  
At the quantum level,
the constraint can be expressed as\footnote{For simple 
illustration, we have omitted
the contribution from the ghost fields.}
\ben
 \bra{\psi^{\prime}}tr\{
\epsilon(t_H)\left(J+\epsilon(K)\right)\}\ket{\psi}=0 
\label{JKc}
\een
Here $\bra{\psi},\bra{\psi^\prime}$ are ghost-free states and $t_H$ 
is any element of the Lie algebra  
of $H$.  
In terms of $J$ and $K$, the conformal stress-tensor
is $T=\frac{1}{2k+c_{G}}:tr_G(JJ):-\frac{1}{2k_{H}+c_{H}}:tr_H(KK):$.
$c_G$ and $c_H$ are the eigenvalues of the quadratic Casimir operators 
in the adjoint representation of $G$ and $H$, respectively. 
Exploiting the constraint, we have $\bra{\psi^{\prime}}T(z)\ket{\psi}=
\bra{\psi^{\prime}}T^G(z)-T^H(z)\ket{\psi}$
with $T^G=\frac{1}{2k+c_{G}}:tr_G(JJ):$ and 
$T^H=\frac{1}{2k+c_{\epsilon(H)}}
:tr_G\{\epsilon(K)\epsilon(K)\}:$. 
This is a field-theoretic version of the GKO stress-tensor, 
confirming that the current $K$ is related to
the current of the Kac-Moody subalgebra via the constraint \cite{kps}.

In order to describe maximally symmetric D-branes in the gauged WZW model,
we will henceforth consider world sheets $\Sigma$ with nonvanishing 
boundary, $\pa\Sigma\neq0$. 
We propose the gluing conditions
\ben
 J|_{\pa \Sigma}=-R\bar{J}|_{\pa \Sigma},\ \ \ \ K|_{\pa\Sigma}
  =-R\bar{K}|_{\pa \Sigma}
\label{R}
\een
with $R$ an automorphism of the Lie algebra of $G$ reducing to $H$. 
When $g=1$, $\gt$ reduces 
to $h$, so the gluing condition $K|_{\pa\Sigma}=-R\bar{K}|_{\pa \Sigma}$ 
can be induced from $J|_{\pa \Sigma}=-R\bar{J}|_{\pa \Sigma}$.
Initially, we shall focus on $R=1$:
\ben
 J|_{\pa\Sigma}=-\bar{J}|_{\pa \Sigma},\ \ \ \ K|_{\pa\Sigma}
  =-\bar{K}|_{\pa \Sigma}
\label{JJb}
\een
In that case, we see that the gluing conditions may be
recast into
\bea
 \gt^{-1}\pa_\tau \gt&=&\frac{1+Ad(\gt)}{1-Ad(\gt)}\gt^{-1}\pa_\sigma\gt\nn
 h^{-1}\pa_\tau h&=& \frac{1+Ad(h)}{1-Ad(h)}h^{-1}\pa_\sigma h
\label{ad}
\eea
with $Ad(g)y=gyg^{-1}$. The coordinates are related by  
$\pa_z=\pa_\tau+\pa_\sigma,\ \pa_{\zb}=\pa_\tau-\pa_\sigma$.
The expressions (\ref{ad}) are sensible because,  
as in the WZW model \cite{AS}, when $\gt$ and
$h$ are restricted to conjugacy classes of $G$ and $H$, respectively, 
the operators $(1-Ad(\gt))$ and $(1-Ad(h))$ are invertible 
when acting on $\gt^{-1}\pa_\sigma\gt$ and 
$h^{-1}\pa_\sigma h$. 
On the boundary, we may therefore choose the parametrizations
\bea
 &&\gt(\tau)=(U^{-1}n)f(U^{-1}n)^{-1}(\tau),\ \ \ \ n,f\in G\nn
 &&h(\tau)=(U^{-1}p)l^{-1}(U^{-1}p)^{-1}(\tau),\ \ \ \ p,l\in H
\label{ghu}
\eea
with $U\in H$. We choose this parametrization so that 
the boundary value of $g$ in (\ref{gb}) below agrees with that in \cite{ES},
up to the change in notation replacing their $k$ with $n$.
Exploiting (\ref{gth}), we then arrive at 
\ben
 g(\tau)=nfn^{-1}plp^{-1}(\tau)
\label{gb}
\een
and
\ben
\ut(\tau)=pl^{-1}p^{-1}U(\tau)
\label{uu}
\een
Thus, the boundary condition (\ref{gb}) for $g$ derived from our gluing
conditions (\ref{JJb}) agrees exactly with that of \cite{Gaw2,ES}:
$g$ is restricted to a product of two conjugacy classes, one for $G$
and one for $H$.
Furthermore, and as expected, we see that the fields $U$ and
$\ut$ are related on the boundary (\ref{uu}).\footnote{If we choose to 
write expressions for the fields in terms of $\ut$, instead of $U$, 
we find that $g$ is a product of elements of conjugacy classes of 
$H$ and $G$, instead of the reverse order: $G$ and $H$. The two conjugacy 
classes are identical in the two cases, however. This is important for the 
Cardy correspondence between boundary conditions and bulk primary fields, 
as we'll see below.} 

Now, since all elements of $G$ and $H$ are conjugate to such  
elements, we can put 
$f=e^{2\pi i\la_{G}/k}$ and $l=e^{2\pi i\la_{H}/k_H}$ 
in (\ref{gb}), where
$\la_{G}$ and $\la_{H}$ are elements of the Cartan subalgebras 
of the Lie algebras of $G$ and $H$, and the factors 
$2\pi/k,2\pi/k_H$ are included 
for later convenience. Extended Weyl invariance implies that we can 
restrict $\lambda_G$ to \cite{Gaw1,ssi} 
\ben
\bar P^k(G)\ =\ \{\lambda_G\,|\, 0\leq \alpha(\lambda)\leq k,\ \forall 
\alpha\in R_>(G)\} 
\label{bPG}
\een
where $R_>(G)$ denotes the set of 
positive coroots of ${\rm Lie}\, G$. 
Similarly, we can restrict to $\lambda_H\in \bar P^{k_H}(H)$. 

Let us now consider the twisted gluing conditions
\ben
 J=-Ad(r)\bar{J},\ \ \ \ K=-Ad(r)\bar{K}
\label{r}
\een
where $r$ is a fixed but arbitrary element of $H$. 
$R$ is thus chosen to be an inner automorphism of 
the Lie algebra of $G$ reducing to $H$ (\ref{R}).
In this case, the elements $\gt$ and $h$ are constrained to {\em twisted}
conjugacy classes of $G$ and $H$ at the boundary, respectively, 
and may be parametrized as
\bea
 \gt(\tau)&=&(U^{-1}n)f(U^{-1}n)^{-1}(\tau)r\nn
 h(\tau)&=&(U^{-1}p)l^{-1}(U^{-1}p)^{-1}(\tau)r
\label{ghur}
\eea
As before, these  boundary conditions are easily translated into those on
$g,U,\ut$:
\ben
 g(\tau)=nfn^{-1}plp^{-1}(\tau)
\label{gbr}
\een 
and
\ben
 \ut(\tau)=pl^{-1}p^{-1}U(\tau)r
\label{uur}
\een
We see that twisting the gluing conditions by an inner automorphism
of the Lie algebra of 
$H$ does not affect the boundary conditions on $g$, even 
though the boundary conditions in each of the sectors (represented
by $\gt,h$) {\it are} affected (\ref{ghur}). The boundary value of $\ut$ 
depends on the automorphism through (\ref{uur}).

\section{Gauged WZW action for open strings}

Let us turn to the gauged WZW action for open strings.
The WZW term is not well-defined for a worldsheet $\Sigma$ with a boundary.
The remedy is to introduce an auxiliary disc $D$ for each hole in $\Sigma$
with boundaries common with those of $\Sigma$. For simplicity,
we consider the situation with a single hole.
The map $g$ from $\Sigma$ to $G$ is then 
extended\footnote{By a common abuse of notation, we use the same symbols 
to denote such extensions as well as the original maps.} to a map from 
the extended worldsheet $\Sigma\cup D$. The disc $D$ is mapped into
the conjugacy classes for $\gt$ and $h$ on the boundary. Following the 
strategy in \cite{AS} and exploiting our gluing conditions, we propose 
\bea
 S^{G/H}&=&\frac{k}{4\pi}\left(\int_\Sigma L^{\rm kin}(\gt)
  +\int_{B}\chi(\gt)-\int_D \om (\gt)\right)\nn
 &-&\frac{k_H}{4\pi}\left(\int_\Sigma L^{\rm kin}(h)
  +\int_{B}\chi(h)-\int_D \om (h)\right)
\label{Sopen}
\eea
as the gauged WZW action for open strings. 
$B$ is a three-dimensional manifold bounded by $\Sigma\cup D$.
$\om (\gt)$ and $\om (h)$ are the 
Alekseev-Schomerus two-forms defined on the conjugacy 
classes of $G$ and $H$, respectively. With our parametrizations, they 
become 
\bea
 \om (\gt)&=&tr\left\{(U^{-1}n)^{-1}d(U^{-1}n)f(U^{-1}n)^{-1}d(U^{-1}n)f^{-1}
  \right\}\nn
 \om (h)&=&tr\left\{(U^{-1}p)^{-1}d(U^{-1}p)l^{-1}(U^{-1}p)^{-1}d(U^{-1}p)l
  \right\}
\label{ww}
\eea
On the conjugacy classes, the three-forms $\chi$ satisfy
\ben
 d\om (\gt)=\chi(\gt),\ \ \ \ d\om (h)=\chi(h)
\label{dw}
\een

The action (\ref{Sopen}) is ambiguous, since one could choose a different 
auxiliary disk $D'$, while keeping the same boundary conditions on 
$\partial\Sigma$. Topologically, $D\cup(-D') \cong S^2$, and so 
the difference in (\ref{Sopen}) with the two choices is related to 
embeddings of $S^2$ into  $G$ and $H$, subject to the boundary conditions 
on $\tilde g$ and $h$, respectively. As explained in \cite{Gaw1,ES}, these 
latter are characterized by coroot lattice vectors $s_G$ of $G$, and 
$s_H$ of $H$, respectively. The corresponding ambiguities in the action 
(\ref{Sopen}) are then  \cite{Gaw1,ES} 
\ben
 \D_{G}S^{G/H}=\frac{k}{4\pi}\left(\int_B \chi (\gt) - 
 \int_{S^{2}}\om (\gt)\right) = 2\pi s_{G}(\lambda_{G})
\label{delta}
\een
and 
\ben
 \D_{H}S^{G/H}=\frac{k_{H}}{4\pi}\left(\int_B \chi (h) - 
 \int_{S^{2}}\om (h)\right) = 2\pi s_{H}(\lambda_{H})
\een
Single-valuedness of path integrals involving the action (\ref{Sopen}) 
therefore leads to
\ben
 \al_{G}(\lambda_{G}) \in  \Z
\label{qg}
\een
\ben
 \al_{H}(\lambda_{H}) \in  \Z
\label{qh}
\een
for any coroots  $\al_{G}$ and $\al_{H}$ of the Lie algebras of  $G$ and $H$. 
Therefore $\lambda_{G}$ and $\lambda_{H}$ are quantized to be 
associated with integral 
weights of those Lie algebras. Furthermore, the allowed set 
\ben
 P^k(G)\ =\ \{\lambda_G\,|\, \alpha(\lambda_G)\in 
  \{0,1,\ldots,k\},\ \forall \alpha\in R_>(G)\} 
\label{PG}
\een
labels the set of bulk primary fields in the WZW model, and similarly 
for $\lambda_H$ in $P^{k_H}(H)$. 

Now, pairs of such Cartan subalgebra elements (and so their 
associated weights)  
$\{\lambda_G,\lambda_H\}$ label the bulk coset primary fields. In the simplest 
cases, this verifies Cardy's assertion that the boundary conditions are 
in one-to-one correspondence with bulk primary fields. When $G$ and $H$ 
share central elements, however, there are the complications of 
selection rules and their dual field identifications, and there is a 
so-called fixed point problem in some cosets. All these were discussed in 
\cite{Gaw2,ES}. Remarkably, Cardy's correspondence continues to 
hold. Possible exceptions are the so-called maverick cosets \cite{DuJo}, 
that are problematic even in the closed string case.

In summary, the gluing condition $J=-\bar J$ ($K=-\bar K$) implies 
that $\tilde g=U^{-1}g\tilde U$ ($h=U^{-1}\tilde U$) is restricted to a 
conjugacy class of the group $G$ (subgroup $H$), labelled by 
$\lambda_{G}\in \bar P^k(G)$ ($\lambda_{H}\in \bar P^{k_H}(H)$). 
The constraint (\ref{JKc}) then 
indicates that $g$ is restricted to be a product of elements labelled by 
$\lambda_{G}$ and $\lambda_{H}$. The quantization conditions 
(\ref{qg}) and (\ref{qh}) then show that 
we can label the boundary states in the coset theory 
by $\{\lambda_{G}, \lambda_{H}\}$ with 
$\lambda_{G}\in P^k(G)$ and  $\lambda_{H}\in P^{k_H}(H)$. 
Our gluing conditions $J=-\bar J$ and $K=-\bar K$,  
with the quantization conditions (\ref{qg}) and (\ref{qh}), 
verify Cardy's correspondence. 

The local gauge transformation
\ben
 g\rightarrow mgm^{-1},\ \ \ \ U\rightarrow mU,\ \ \ \ \ut\rightarrow m\ut
\label{gauge}
\een
with $m\in H$, reduces on the boundary to
\ben
 n\rightarrow mn,\ \ \ \ p\rightarrow mp,\ \ \ \ U\rightarrow mU,\ \ \ \ 
  \ut=pl^{-1}p^{-1}U\rightarrow mpl^{-1}p^{-1}U
\label{gauge2}
\een
Since $\om (\gt)$ and $\om (h)$ are invariant under the (reduced)
local gauge transformations
(\ref{gauge2}), the gauged WZW action for open strings (\ref{Sopen})
is invariant under the transformation (\ref{gauge}).

We may rewrite the gauged WZW action (\ref{Sopen}) in terms of the gauge
fields $A$ and the related $H$-valued fields $U,\ut$. 
Classically, i.e., neglecting possible 
quantum corrections from the path-integral
measure, we arrive at
\bea
 S^{G/H}&=&S^G(g)+\frac{k}{2\pi}\int_\Sigma d^2z\,tr\left\{
  A_{\zb}\pa_zgg^{-1}-A_zg^{-1}\pa_{\zb}g+A_{\zb}gA_z g^{-1}-A_z A_{\zb}
  \right\}\nn
 &+&\frac{k}{4\pi}\int_D d^2z\,{\cal{C}}
\label{SC}
\eea
with
\bea
\cal {C}&=&-\om(\gt)+\om(h)-tr\left\{ g^{-1}dgd\ut\ut^{-1}
  -dUU^{-1}(dgg^{-1}+gd\ut\ut^{-1}g^{-1})\right\}\nn
 &+&dUU^{-1}d\ut\ut^{-1}
\label{C}
\eea
Since the images of $D$ are also constrained to the conjugacy classes,
we extend the boundary conditions (\ref{gb}) and (\ref{uu})
to $D$. $\cal C$ is defined on $D$ only, so after straightforward
calculation, it can be rewritten as
\ben
 {\cal{C}} =-\left(\om(n)+\om(p)+tr(dc_2c_2^{-1}c_1^{-1}dc_1)\right)
\label{C2}
\een
Here $c_1=nfn^{-1}$ and $c_2=plp^{-1}$, elements of conjugacy classes
of $G$ and $H$, respectively, and 
\ben
 \om(n)=tr(n^{-1}dnfn^{-1}dnf^{-1}),\ \ \ \ 
  \om (p)=tr(p^{-1}dplp^{-1}dpl^{-1})
\label{ww2}
\een 
Notice that $\om (n)$ and $\om (p)$ just written are different from 
$\om (\gt)$ and $\om (h)$ given in (\ref{ww}) above.
Eqs. (\ref{C2}) and (\ref{ww2}) show that all the $U$-dependent terms
in (\ref{C}) cancel in a nontrivial way.
Inserting (\ref{C2}) in the action (\ref{SC}), we recover the action of
\cite{Gaw2,ES} (see (2.9) of \cite{Gaw2} and (3.11) of \cite{ES}). 
Thus, the open string action (\ref{Sopen}) constructed
from our gluing conditions can be reduced classically 
to those written previously in a different approach.

To conclude this section, let us discuss the 
relation of the gluing 
conditions to   
the currents 
\bea
 &&j=-\pa_z gg^{-1}-gA_zg^{-1}+A_z\nn
 &&\bar{j}=g^{-1}\pa_{\zb}g-g^{-1}A_{\zb}g+A_{\zb}
\label{jjb2}
\eea
defined by varying 
the gauge fields $A$ \cite{vdp} in a gauged WZW action.   
One might hope that our gluing conditions could be expressed 
in a natural way  
using $j,\bar j$. This does not turn out to be 
the case, however. The only simple relations between the 
currents $j,\bar j$ and $J,\bar J,K,\bar K$ are
\bea
 &&U^{-1}jU\ =\ J+K\nn
 &&\ut^{-1}\bar j\ut\ =\ \bar J +\bar K
\label{jJu}
\eea
and they also involve the fields $U,\ut$. 
The complete set of gluing conditions 
(\ref{JJb}) cannot be rewritten in terms of $j,\bar j$, only, and so 
D-branes in the gauged
WZW model cannot be characterized in that manner.  
While the gluing conditions can be rewritten 
in terms of $j,\bar j$ and $U,\ut$, their form is not enlightening.

\section{Conclusion}

We have proposed the gluing conditions $J=-\bar J$ and $K=-\bar K$ 
to describe D-branes in the gauged WZW model. The gauged WZW action
can be written as  $S^{G/H}=S^G(\gt)-S^H(h)$, showing that
the fields $\gt,h$ formally decouple  --
we can therefore introduce two currents $J$ and $K$ for the group 
$G$ and the subgroup $H$.
The boundary conditions for $g,U,\ut$ have been derived
from the gluing conditions, and the boundary condition on $g$
proposed in \cite{Gaw2,ES} has been recovered.

We have also considered gluing conditions twisted by inner automorphisms  
of $G$ reducing to $H$. It was found that in the product of
the twisted conjugacy classes, the twistings cancel, and we are
left with the same product of (untwisted) conjugacy classes as in the 
untwisted case.

From our gluing conditions, 
we have constructed the gauged WZW action for 
open strings, with the new feature that the two-forms $\om$ depend on the
boundary value of one of the $H$-valued fields $U,\ut$.
When rewriting our action in terms of the gauge field $A$, we have 
shown that all the $U$-dependent terms cancel each other 
in a nontrivial way, and the resulting action 
coincides classically with that in \cite{Gaw2,ES}. This indicates 
that these gluing conditions indeed describe D-branes in the gauged WZW model.

\vskip.5cm
\noindent{\bf Acknowledgements}
\vskip.1cm
\noindent 
The work of T.K. is supported in part by a grant from the Ministry of 
Education (grant number 13135215). J.R. is supported by a 
CRM-ISM postdoctoral fellowship.
The research of M.A.W. and J.-G.Z. is supported in part by NSERC.

\end{document}